\documentclass[final,5p,times,twocolumn]{elsarticle}

\usepackage{amssymb}
\usepackage{color}
\usepackage{amsmath}
\usepackage{multirow}
\usepackage{booktabs}
\usepackage{bm}
\usepackage{braket}
\usepackage{amsmath}
\usepackage{graphicx}
\usepackage{soul}
\usepackage {ulem}
\biboptions{sort&compress}
\usepackage{subcaption}
\usepackage{tabularx}

\usepackage[pdfstartview=FitH, colorlinks,
 linkcolor={red!60!black!},
 anchorcolor={green!80!black!},
 urlcolor={blue!80!black!},
 citecolor={blue!50!black!}]{hyperref}
\usepackage[table]{xcolor}

\journal{Physics Letters B}
\newcolumntype{Y}{>{\centering\arraybackslash}X}

\begin{document}

%%\begin{CJK}{UTF8}{gbsn} %Baishan Hu

\begin{frontmatter}



%\title{ {\it Ab initio} perturbative valence-space Hamiltonians for the Gamow shell model calculations}

\title{Mechanisms of mirror energy difference for states exhibiting Thomas-Ehrman shift: Gamow shell model case studies of $^{18}$Ne/$^{18}$O and $^{19}$Na/$^{19}$O}

\author[ad1,ad2,ad3]{J.G. Li\corref{correspondence}}
\author[ad4,ad1]{K. H. Li}
\author[ad1,ad2,ad3]{N. Michel\corref{correspondence}}
\author[ad1,ad2]{H. H. Li}
\author[ad1,ad2,ad3]{W. Zuo\corref{correspondence}}

\address[ad1]{CAS Key Laboratory of High Precision Nuclear Spectroscopy, Institute of Modern Physics,
Chinese Academy of Sciences, Lanzhou 730000, China}
\address[ad2]{School of Nuclear Science and Technology, University of Chinese Academy of Sciences, Beijing 100049, China}
\address[ad3]{Southern Center for Nuclear-Science Theory (SCNT), Institute of Modern Physics, Chinese Academy of Sciences, Huizhou 516000, Guangdong Province, China}

\address[ad4]{Institute of Particle and Nuclear Physics, Henan Normal University, Xinxiang 453007, China}

\cortext[correspondence]{Corresponding author:\\ jianguo\_li@impcas.ac.cn(J.G. Li), nicolas.michel@impcas.ac.cn(N. Michel), zuowei@impcas.ac.cn(W. Zuo)}

\begin{abstract}

The mirror energy difference (MED) of the mirror state, especially for states bearing the Thomas-Erhman shift, serves as a sensitive probe of isospin symmetry breaking. We employ the Gamow shell model, which includes the inter-nucleon correlation and continuum coupling, to investigate the MED for $sd$-shell nuclei by taking the $^{18}$Ne/$^{18}$O and $^{19}$Na/$^{19}$O as examples. Our GSM provides good descriptions for the excitation energies and MEDs for the $^{18}$Ne/$^{18}$O and $^{19}$Na/$^{19}$O. Moreover, our calculations also reveal that the large MED of the mirror states is caused by the significant occupation of the weakly bound or unbound  $s_{1/2}$ waves, giving the radial density distribution of the state in the proton-rich nucleus more extended than that of mirror states in deeply-bound neutron-rich nuclei. Furthermore, our GSM calculation shows that the contribution of Coulomb is different for the low-lying states in proton-rich nuclei, which significantly contributes to MEDs of mirror states. Moreover, the contributions of the nucleon-nucleon interaction are different for the mirror state, especially for the state of proton-rich nuclei bearing the Thomas-Erhman shift, which also contributes to the significant isospin symmetry breaking with large MED.

\end{abstract}

\end{frontmatter}

\textit{Introduction.}~
Exotic nuclear structures in drip-line nuclei have become a subject of great interest in recent years, as they are characterized by unique properties that exhibit significant differences compared to those of stable nuclei.
One of the most significant phenomena observed in these systems is the Thomas-Ehrman shift (TES)~\cite{PhysRev.88.1109, PhysRev.81.412}. This effect is most pronounced in nuclei close to the proton drip lines, where the balance between the strong force and the Coulomb force is most delicate. 
States exhibiting TES effects are often weakly bound or unbound, characteristic of open quantum systems, while their neutron-rich mirror counterparts remain deeply bound, resulting in a large mirror energy difference (MED) in their isobaric states~\cite{PhysRev.88.1109, PhysRev.81.412,PhysRevLett.102.152502,PhysRevLett.125.192503,Sun_2024,PhysRevC.107.014302}. 
The large MEDs are attributed to their proximity to near-threshold effects, in which the continuum effects need to be well treated.
A thorough comprehension of the Thomas-Ehrman shift is pivotal for elucidating the dynamics of weakly bound and unbound nuclear systems and understanding the mechanisms underlying isospin symmetry breaking in mirror nuclei.

Two possible reasons exist for the states with large MED, of external or internal character.
If extended single-particle wave functions of weakly- or unbound $s$- or $p$-waves are significantly occupied in the considered states, the large MED is of external nature, as in the TES states \cite{PhysRevC.89.044327,PhysRevLett.125.192503,PhysRevC.90.014307}.
The second possibility, related to configuration mixing (see Refs.\cite{LI2022137225,PhysRevC.102.024309}), is of internal nature. In this case, the extended wave function is given via the strong configuration mixing in which a few nodal states of $s$ or $p$ waves are included in the calculations.
These two external and internal effects are different but can be intertwined in a complex manner.
For instance, the inversion of ground states in the $^{16}$F and $^{16}$N mirror nuclei is primarily due to the unbound proton $1s_{1/2}$ orbital, which can also be well-described in GSM calculation within the configuration mixing framework~\cite{PhysRevC.90.014307,PhysRevC.106.L011301}. 

The $sd$-shell nuclei, situated at the boundary between the light and heavy nuclei, exhibit a wide range of nuclear structure phenomena that remain somewhat mysterious \cite{doi:10.1126/science.122.3170.603.b}.
In recent years, these nuclei have been extensively studied using a variety of experimental techniques \cite{RevModPhys.77.427, AJZENBERGSELOVE19881, CAMPBELL2016127}.
%These studies have not only permitted a more detailed exploration of their unique nuclear structures but also resulted in more accurate and precise measurements of MEDs, thereby shedding light on ISB \cite{CAMPBELL2016127}.
A wealth of information on the Thomas-Ehrman shift has been gleaned from $sd$-shell proton drip-line nuclei, where numerous states exhibiting significant TES effects have been identified~\cite{PhysRevLett.102.152502,PhysRevC.67.014308,PhysRevLett.125.192503,Sun_2024}.
For instance, the mirror pairs $^{18}$Ne/$^{18}$O~\cite{PhysRevLett.102.152502,ZHANG2022136958} and $^{19}$Na/$^{19}$O~\cite{PhysRevC.67.014308} serve as notable examples.
For the $sd$-shell nuclei, TES is mainly driven by $s$-waves. Indeed, the proton $1s_{1/2}$ orbital is weakly bound or unbound in proton drip-line nuclei, whereas the neutron $1s_{1/2}$ is well-bound in their mirror neutron-rich nuclei.

Several theoretical models have been developed to probe the isospin asymmetry for mirror nuclei, especially the MEDs of $sd$-shell nuclei, such as the standard shell model (SM)~\cite{BENTLEY2007497, PhysRevLett.89.142502, PhysRevLett.110.172505, PhysRevC.87.054304}, mean-field calculations \cite{BACZYK2018178,PhysRevLett.124.152501}, and \textit{ab initio} approaches \cite{PhysRevC.104.024319, PhysRevC.104.014324, PhysRevC.66.024314, ZHANG2022136958,LI2023138197}.Within the standard SM calculations, weakly bound and unbound wave functions on eigenenergies are indirectly considered via phenomenologically adjusting the matrix element related to $1s_{1/2}$ orbit~\cite{PhysRevC.89.044327,PhysRevLett.125.192503}. 
Mean-field calculations, such as the Skyrme-Hartree-Fock, have also been extensively employed in MED studies~\cite{BACZYK2018178,PhysRevLett.124.152501}. However, these models involve parameters that are constrained by data \cite{BENTLEY2007497, BACZYK2018178, PhysRevC.106.024327}. In recent years, \textit{ab initio} approaches, such as the \textit{ab initio} valence-space in-medium similarity renormalization group have also been applied to study MEDs of $sd$-shell nuclei~\cite{PhysRevC.104.024319, PhysRevC.104.014324, PhysRevC.66.024314, ZHANG2022136958,PhysRevC.107.014302,Sun_2024,LI2023138197,CPC:10.1088/1674-1137/acf035}, in which, the extended many-body wavefunctions are partially described by using a large number of HO spaces. 
Moreover, current theoretical calculations have pointed out that the TES is caused by the repulsive Coulomb interaction and the occupations of weakly bound or unbound $s$- or $p$-waves for valence protons.
However, detailed studies on the mechanism of the TES are also lacking.

One of the major challenges in studying drip line nuclei is accounting for the interplay between configuration-mixing and continuum effects. The Gamow shell model (GSM)~\cite{PhysRevLett.89.042501,PhysRevLett.89.042502,0954-3899-36-1-013101,Michel_GSM_book,PhysRevC.102.034302,ZHANG2022136958,PhysRevC.104.L061306,LI2022137225,PhysRevC.104.024319,PhysRevC.102.024309} has emerged as a powerful tool in this regard, as it provides a unified framework to describe the structure of nuclei close to the particle emission threshold and allows for a accurate understanding of the exotic properties in drip line nuclei. Based on the above situation, we employ the GSM to investigate the significant isospin symmetry breaking with large MED values and behind mechanism for the $sd$-shell nuclei, taking the $^{18}$Ne/$^{18}$O and $^{19}$Na/$^{19}$O mirror partner as examples.

\textit{Method.---}~
GSM is a multiconfiguration shell model framework, which works in the picture of a core plus valence nucleons~\cite{0954-3899-36-1-013101,Michel_Springer,physics3040062,PhysRevC.103.034305,PhysRevC.108.064316}. 
At the heart of GSM lies the utilization of the one-body Berggren basis~\cite{BERGGREN1968265}, which possesses bound, resonance, and scattering states, generated by a finite-range potential, typically of Woods-Saxon (WS) types (see details in Ref.~\cite{BERGGREN1968265,0954-3899-36-1-013101,Michel_Springer}). 
%The completeness of the Berggren basis for a given partial wave characterized by quantum numbers $\ell,j$ is demonstrated by the following relation:
%\begin{equation}
%\sum_n u_{n}^{(\ell j)}(r) u_{n}^{(\ell j)}(r') + \int_{L^+} u_{k}^{(\ell j)}(r)  u_{k}^{(\ell j)}(r') ~dk = \delta(r - r'), \label{Berggren}
%\end{equation}
%where $n$ enumerates the bound and resonance states of the considered partial wave, while $L^+$ is the complex contour of scattering states, which encompasses the resonance states present in the discrete sum.
%For practical numerical applications of Eq.(\ref{Berggren}), the contour of integration requires discretization. This is efficiently accomplished through the Gauss-Legendre quadrature~\cite{PhysRevC.83.034325}. %This procedure yields a discretized version of the Berggren completeness relation, effectively mirroring the completeness relation of the harmonic oscillator basis.
Contrary to the traditional SM, the GSM Hamiltonian matrix is characterized as complex symmetric~\cite{0954-3899-36-1-013101,Michel_Springer}.
Moreover, the GSM Hamiltonian possesses numerous many-body scattering states, so that the bound or resonance eigenstates are embedded among scattering eigenstates. Thus, one developed the overlap method along with the Jacobi-Davidson method extended to complex-symmetric matrices to diagonalize and identify many-body resonance eigenstates~\cite{0954-3899-36-1-013101,MICHEL2020106978,Michel_Springer}. In the case of well-bound states in traditional SM calculations, the use of the Lanczos method is optimal~\cite{RevModPhys.77.427}.
As a consequence, the GSM calculation includes both the inter-nucleon correlations and continuum coupling~\cite{0954-3899-36-1-013101,Michel_Springer,physics3040062}.

The many-body Schr{\"o}dinger equation of GSM Hamiltonian can be solved within the so-called cluster orbital shell model (COSM) formalism~\cite{PhysRevC.38.410} (see Refs.~\cite{PhysRevC.84.051304,PhysRevC.96.054316,Michel_Springer}). 
%As the GSM Hamiltonian is defined in the core+valence particle picture, we can solve the many-body Schr{\"o}dinger equation of GSM Hamiltonian within the so-called cluster orbital shell model (COSM) formalism~\cite{PhysRevC.38.410} (see Refs.~\cite{PhysRevC.84.051304,PhysRevC.96.054316,Michel_Springer}).
%Within the COSM framework, one defines the radial coordinates of valence nucleons with respect to the center of mass of the core. The coordinates of valence nucleons are translationally invariant, thereby suppressing all centers of mass excitations by definition. 
The GSM Hamiltonian in COSM coordinates reads~\cite{PhysRevC.84.051304,PhysRevC.96.054316,Michel_Springer}:
\begin{equation}
\!\!\!\hat{H}_{\rm GSM} \!=\!\! \sum_{i=1}^{A_{val}} \left( \frac{\mathbf{p}_i^2}{2 \mu_i} + \hat{U}_i^{(c)} \right) + \sum_{i<j}^{A_{val}} \left(  \hat{V}_{ij} + \frac{\mathbf{p}_i \cdot \mathbf{p}_j}{M_{c}} \right), \label{H_GSM}
\end{equation}
where $A_{val}$ is the number of valence nucleons, $\mu_i$ is the effective mass of the nucleon, $\hat{U}_i^{(c)}$ is represented by a one-body WS potential mimicking
the inert core.
$\hat{V}_{ij}$ is the residual inter-nucleon interaction, which is modeled by a pionless effective field theory (EFT) interaction~\cite{RevModPhys.85.197,RevModPhys.92.025004}, and the last term embodies the recoil effects induced by the finite mass of the core $M_{c}$. 
Concerning the used pionless EFT interaction \cite{RevModPhys.85.197,RevModPhys.92.025004}, only two-body contact terms up to next-to-next leading-order are considered in the present calculations, and EFT interaction is optimized to reproduce the low-lying states of selected nuclei.
The harmonic oscillator (HO) basis used for the representation of the EFT interaction is limited to a few shells.
This regularization approach has been recently utilized in Refs.~\cite{PhysRevC.93.044332,PhysRevC.86.031301,PhysRevC.98.054301,PhysRevC.98.044301}, and in particular in GSM calculation of unbound neutron-rich oxygen isotopes~\cite{PhysRevC.103.034305}.

In the present work, the $^{18}$Ne/$^{18}$O and $^{19}$Na/$^{19}$O mirror partners are taken as examples. The doubly magic nucleus $^{16}$O is chosen as the inert core, and the $s_{1/2}$, $p_{1/2,3/2}$ and $d_{3/2,5/2}$ partial waves are represented by the Berggren basis, in which 40 discretization points are used in total for continuum states in each partial wave, and the $f_{5/2,7/2}$ partial waves are treated using the HO basis, in which 6 HO states are adopted for each partial wave.
Only the Coulomb force is considered for the isospin non-conserving part of the GSM Hamiltonian. The contribution of the isospin-dependent part of nuclear interaction to TES is small, which is neglected in the present GSM calculations.
The adopted Hamiltonian is optimized to produce the selected experimental data of $sd$-shell nuclei. The calculated excitation energies of $^{18}$Ne/$^{18}$O and $^{19}$Na/$^{19}$O mirror partners are presented in Tables. \ref{tab:18Ne-18O-Ex} and \ref{tab:19Na-19O-Ex}, which show good agreements with experimental data~\cite{ensdf}. In the following section, the mechanics of mirror energies difference for the $^{18}$Ne/$^{18}$O and $^{19}$Na/$^{19}$O mirror partners are detail investigated.
\begin{table}[!htb]
    \centering
    \setlength{\tabcolsep}{0.4mm}
    \begin{tabular}{cccccccccc}
    \hline\hline
    \multirow{2}*{$J^\pi$ } &\multicolumn{4}{c}{$^{18}$Ne} & ~ & \multicolumn{2}{c}{$^{18}$O} \\
    \cmidrule{2-5} \cmidrule(r){7-8}
    %\cline{2-5} \cline{7-8}
    & $E_{\rm exp}$ & $E_{\rm GSM}$ & $\Gamma_{\rm exp}$ & $\Gamma_{\rm GSM}$ &~& $E_{\rm exp}$ & $E_{\rm GSM}$ & MED$_{\rm exp}$ & MED$_{\rm GSM}$\\
    \hline
    $0_1^+$ & 0 & 0 & & 0 &~& 0 & 0 & 0 & 0 \\
    $2_1^+$ & 1.89 & 1.83 &  & 0 & ~ & 1.98 & 1.93&$-95$&$-106$\\
    $4_1^+$ & 3.38 & 2.72 &  & 0 & ~& 3.56 & 2.72&$-179$&$-3$\\
    $2_2^+$ & 3.62 & 3.98 &  & 0 & ~& 3.92 & 4.40&$-304$&$-420$\\
    $0_2^+$ & 3.58 & 4.58 &  & 0 & ~&  3.63 & 5.42&$-58$&$-834$\\
    $3_1^+$ & 4.56 & 4.63 & 18 & 80 & ~& 5.38 & 5.53&$-817$&$-892$\\
    
    \hline\hline
    \end{tabular}
    \caption{The calculated excitation energies of $^{18}$Ne/$^{18}$O with GSM calculations, along with experimental data~\cite{ensdf}. The unit of excitation energy is given in MeV, and the units of particle decay width and MED are given in keV.}
    \label{tab:18Ne-18O-Ex}
\end{table}

\begin{table}[!htb]
    \centering
    \setlength{\tabcolsep}{0.6mm}
    \begin{tabular}{ccccccccccc}
    \hline\hline
    \multirow{2}*{$J^\pi$ } &\multicolumn{2}{c}{$^{19}$Na} & ~ & \multicolumn{2}{c}{$^{19}$O} \\
    \cmidrule{2-3} \cmidrule(r){5-6}
    %\cline{2-5} \cline{7-8}
    & $E_{\rm exp}$ & $E_{\rm GSM}$ &~& $E_{\rm exp}$ & $E_{\rm GSM}$& MED$_{\rm exp}$ & MED$_{\rm GSM}$ \\
    \hline
    $5/2_1^+$ & 0 & 0  &~& 0 & 0 &0&0\\
    $3/2_1^+$ & 0.12 & 0.64  & ~ & 0.10 & 0.71&24&$-64$\\
    $1/2_1^+$ & 0.75 & 0.69 & ~& 1.47 & 1.30&$-727$&$-607$\\
    
    \hline\hline
    \end{tabular}
    \caption{Similar to Table \ref{tab:18Ne-18O-Ex}, but for  $^{19}$Na/$^{19}$O.}
    \label{tab:19Na-19O-Ex}
\end{table}

%\clearpage
%\section{Results}
\textit{Results.}---
Our GSM calculations accurately describe the excitation energies of low-lying states for the mirror partners $^{18}$Ne/$^{18}$O and $^{19}$Na/$^{19}$O. To delve deeper into the significant isospin symmetry breaking observed in these mirror nuclei, we define the MED for a given state $J^\pi$ as MED$(J^\pi)$ = $E_x (T_z^<, J^\pi) - E_x (T_z^>, J^\pi)$, where $T_z^{<}$ and $T_z^{>}$ denote the negative and positive isospin projection $T_z = (N-Z)/2$, respectively, for a mirror pair.
We have calculated MED values for mirror states in $^{18}$Ne/$^{18}$O and $^{19}$Na/$^{19}$O, as presented in Tables \ref{tab:18Ne-18O-Ex} and \ref{tab:19Na-19O-Ex}, along with experimental data.
It is observed that the calculated MED values for low-lying states in the mirror partners $^{18}$Ne/$^{18}$O and $^{19}$Na/$^{19}$O, align well with the experimental data. However, an exception is noted for the $0_2^+$ state in the $^{18}$Ne/$^{18}$O mirror nuclei, where our GSM calculations yield larger values than the experimental data.
Both our GSM calculations and the experimental data highlight significant isospin symmetry breaking in the $3_1^+$ state of the $^{18}$Ne/$^{18}$O mirror nuclei and the $1/2_1^+$ state of the $^{19}$Na/$^{19}$O mirror nuclei, evidenced by their large MED values.

\begin{figure}[!htb]
\includegraphics[width=1.00\columnwidth]{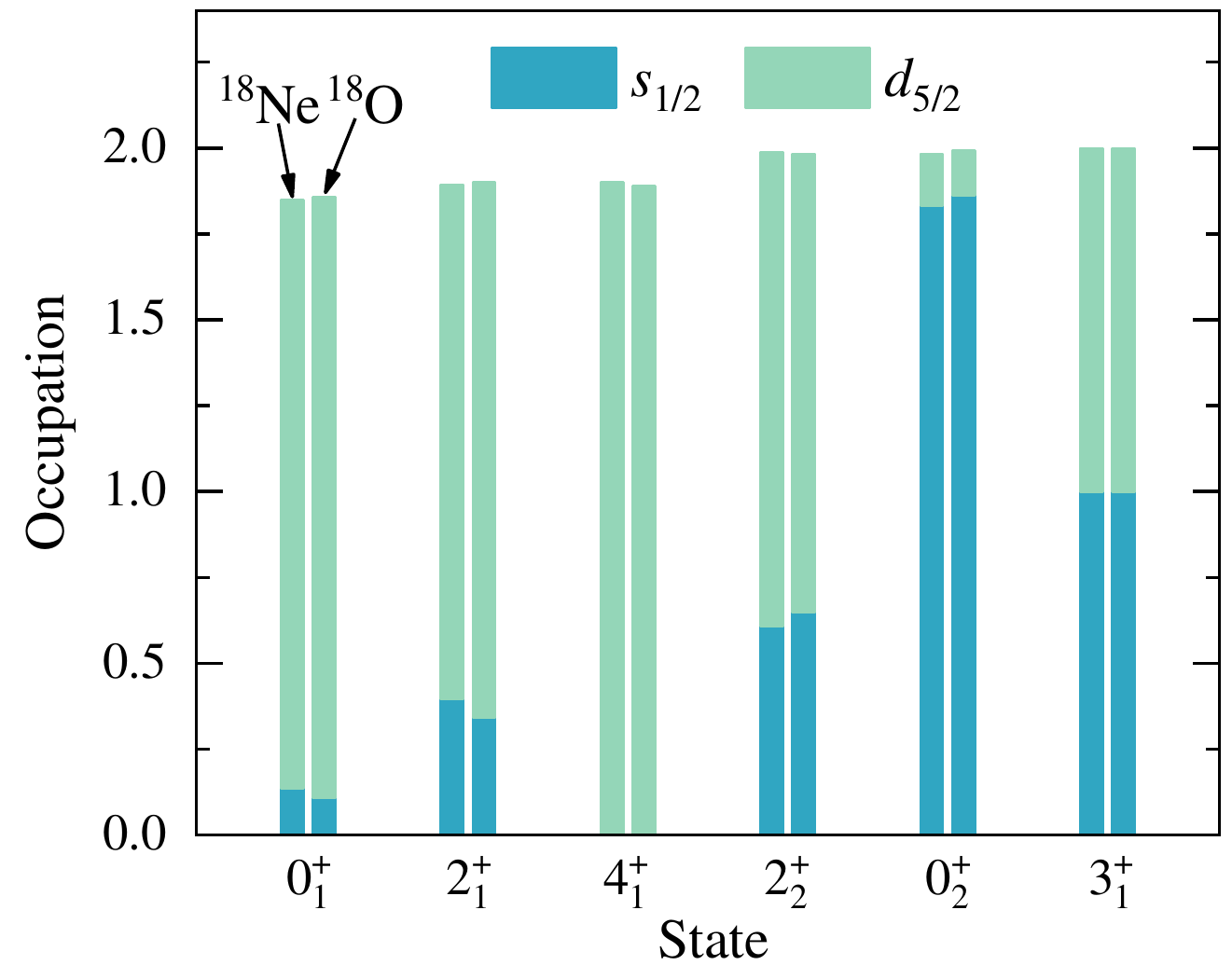}% Here is how to import EPS art
\caption{The average occupation numbers for the $s_{1/2}$ and $d_{5/2}$ partial waves in the low-lying states of the $^{18}$Ne/$^{18}$O mirror pair, calculated using the GSM above the $^{16}$O core.}{}\label{fig-18Ne-18O-occupation}
\end{figure}

\begin{figure}[!htp]
\includegraphics[width=1.00\columnwidth]{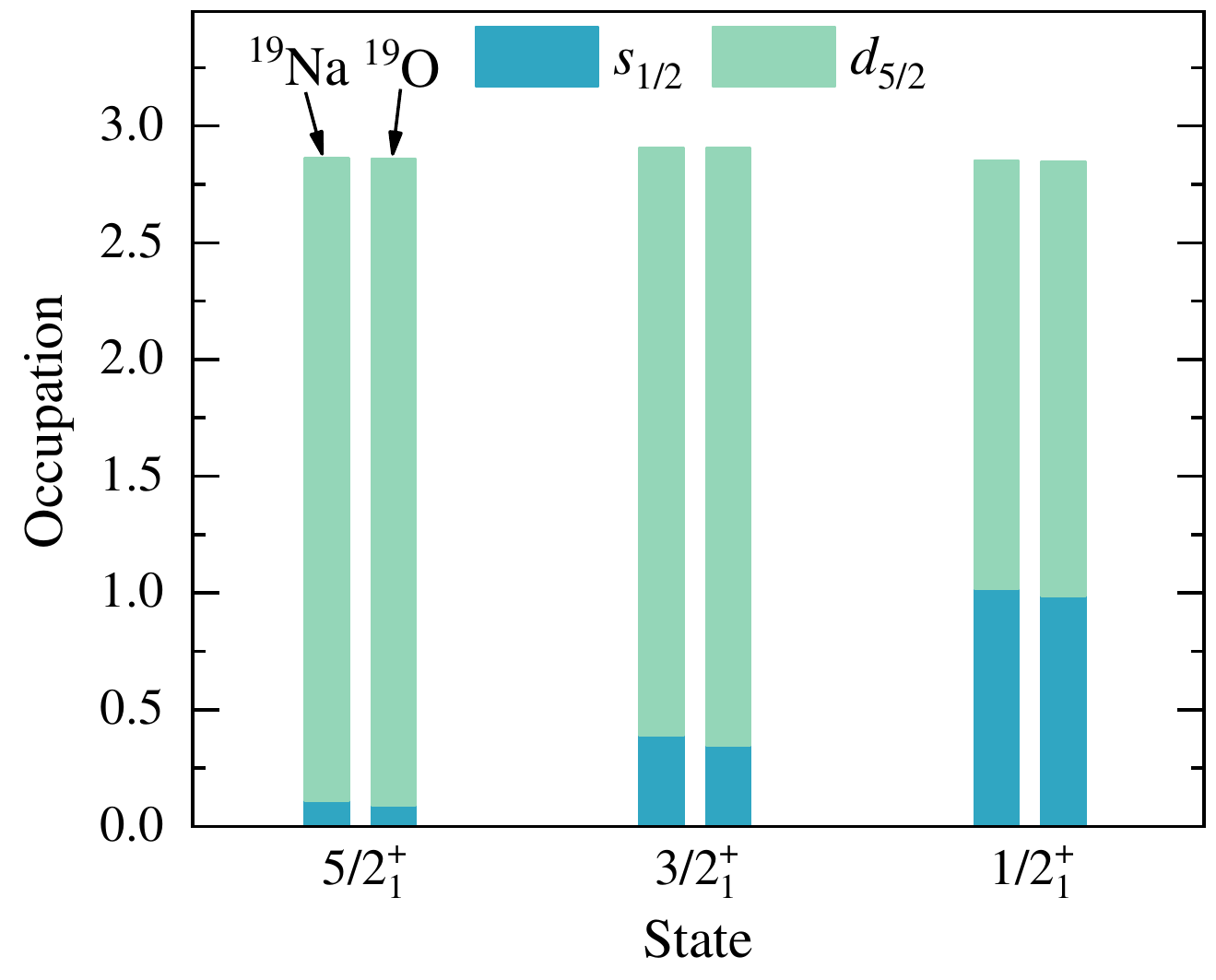}% Here is how to import EPS art
\caption{Similar to Fig.~\ref{fig-18Ne-18O-occupation}, but for low-lying states in $^{19}$Na/$^{19}$O.}{}\label{fig-19Na-19O-occupation}
\end{figure}

To investigate the significant isospin symmetry breaking and the associated large MEDs, we begin with calculating the average occupations of low-lying states through the GSM. The focus is particularly on the $s_{1/2}$ and $d_{5/2}$ partial waves above the $^{16}$O core for the $^{18}$Ne/$^{18}$O and $^{19}$Na/$^{19}$O mirror nuclei, as illustrated in Figs. \ref{fig-18Ne-18O-occupation} and \ref{fig-19Na-19O-occupation}. Notably, other partial waves like $d_{3/2}$, $f_{5/2,7/2}$ exhibit negligible occupations and are, therefore, excluded from these figures.
The calculated average occupations reveal almost identical patterns for mirror states within the $^{18}$Ne/$^{18}$O and $^{19}$Na/$^{19}$O pairs. Our GSM calculations further indicate that states exhibiting significant isospin symmetry breaking with large MED values also show significant occupancy in the $s_{1/2}$ partial waves—markedly higher than in their respective ground states. 
For instance, the occupations of the $s_{1/2}$ partial wave for the $3_1^+$ and $1/2_1^+$ states in the $^{18}$Ne/$^{18}$O and $^{19}$Na/$^{19}$O mirror pairs, respectively, are substantially greater than those of the ground states. 
Additionally, our calculations give that the $0_2^+$ of $^{18}$Ne/$^{18}$O demonstrate a notable $s_{1/2}$ partial wave occupations compared to the ground states, resulting in a large MED.  Contrastingly, experimental data give a smaller MED value, hinting at a complex structure of the $0_2^+$ states in $^{18}$Ne/$^{18}$O that might not be fully captured by  $^{16}$O plus valence particles picture.
%These results are similar to the results from other theoretical models, including the standard shell model and \textit{ab initio} VS-IMSRG calculations~\cite{PhysRevC.107.014302}.

\begin{figure}[!htb]
\includegraphics[width=1.00\columnwidth]{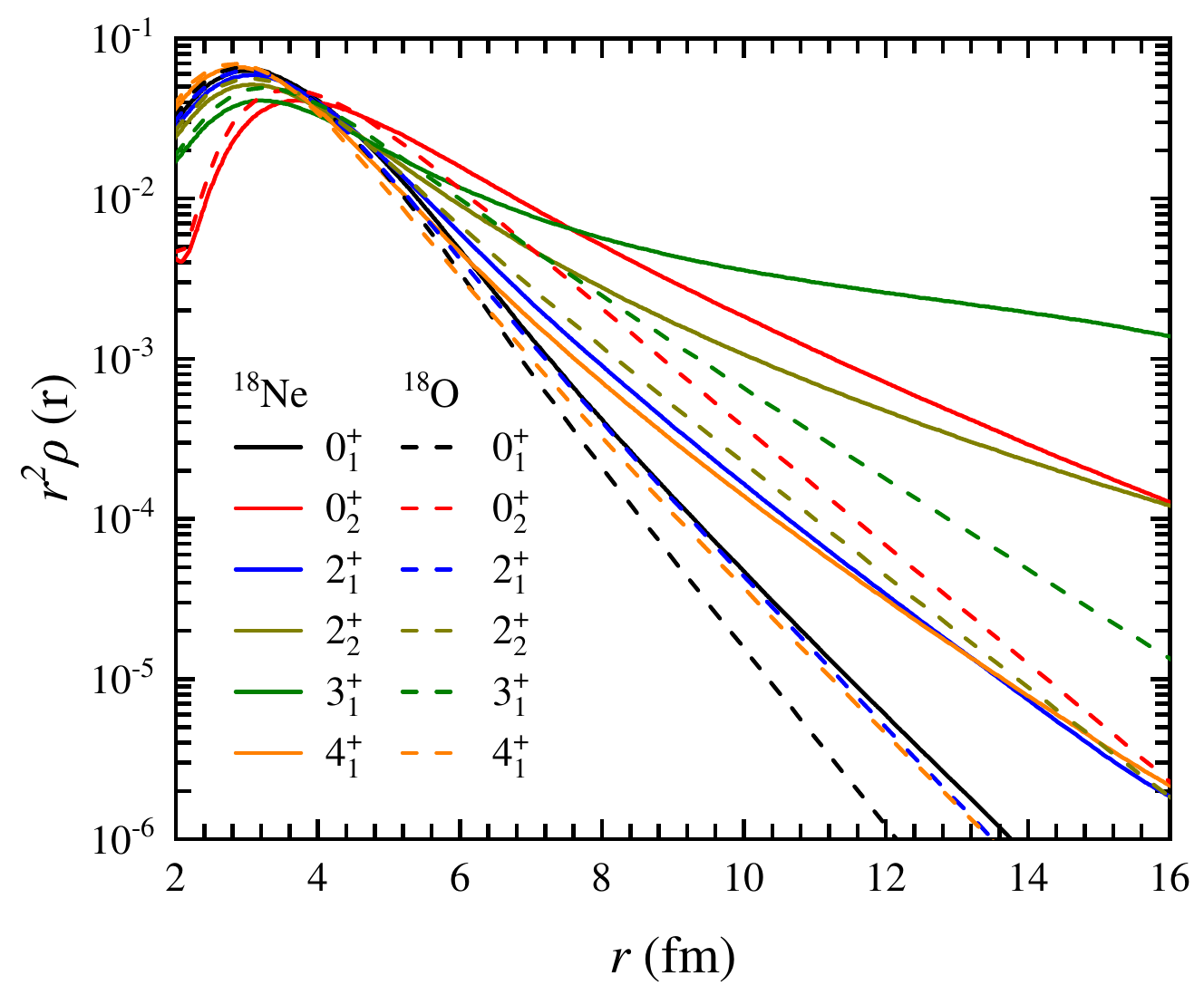}% Here is how to import EPS art
\caption{The calculated radial density distribution of valence protons and valence neutrons for low-lying mirror states in $^{18}$Ne/$^{18}$O, respectively, above $^{16}$O inner core, using GSM.}{}\label{fig-18Ne-18O-density}
\end{figure}

\begin{figure}[]
\includegraphics[width=1.00\columnwidth]{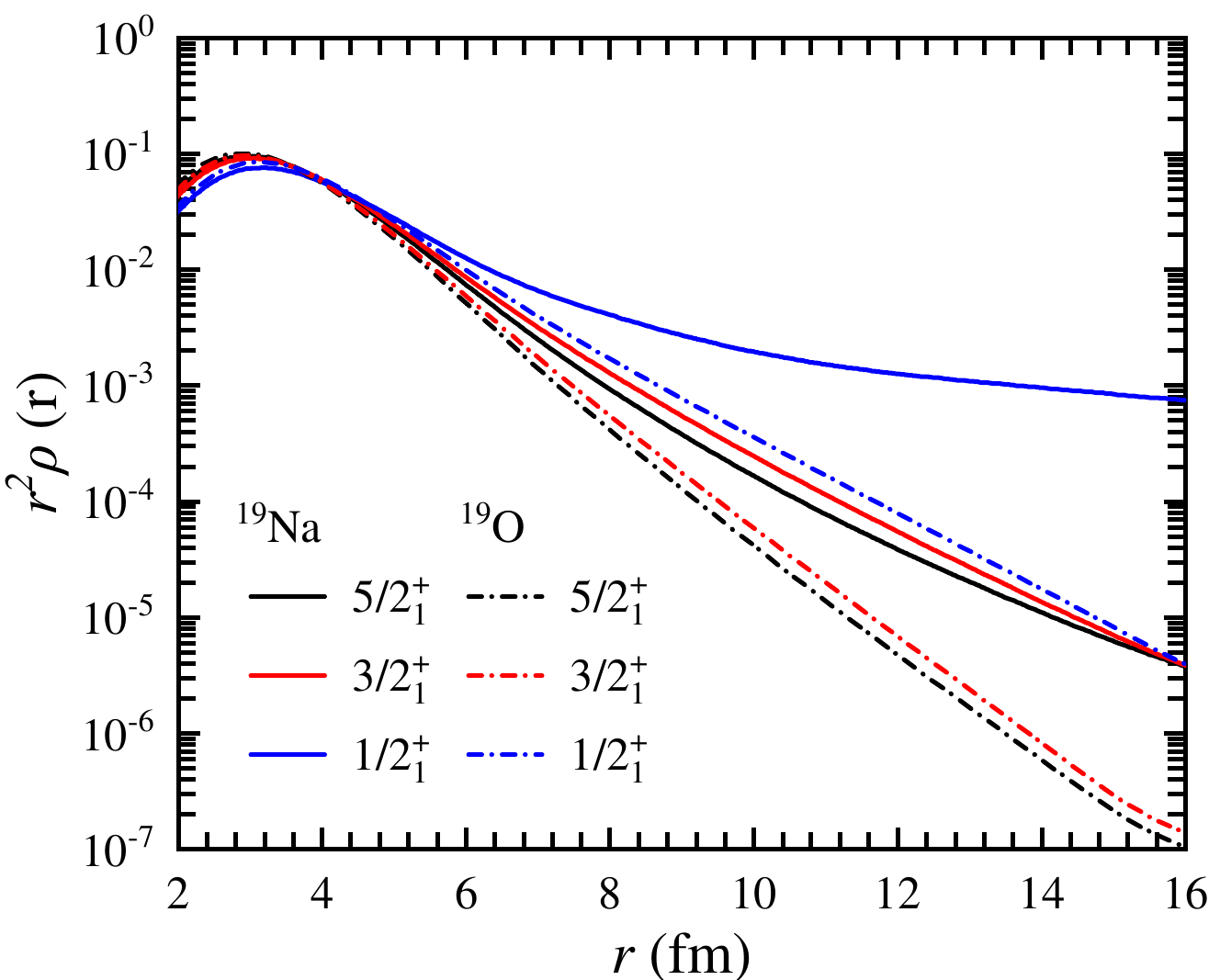}% Here is how to import EPS art
\caption{Similar to Fig.~\ref{fig-18Ne-18O-density}, but for low-lying states in $^{19}$Na/$^{19}$O.}{}\label{fig-19Na-19O-density}
\end{figure}

Aligned with results from other theoretical frameworks, such as the standard shell model and the \textit{ab initio} VS-IMSRG approach, our results indicate that the significant isospin symmetry breaking with large MEDs observed in mirror states stem primarily from the extensive occupation of the $s_{1/2}$ partial waves, which is weakly bound or unbound in the proton-rich nucleus but deeply bound in its mirror neutron-rich nucleus, called TES.
However, a further deep understanding of the mirror state bearing significant isospin symmetry breaking with large MED value is lacking.
The GSM is a very suitable model, which properly treats both the inter-nucleon correlations and continuum coupling, to describe the properties of dripline nuclei, including a precious description of the many-body wave function in the asymptotic regions \cite{PhysRevC.103.034305,PhysRevC.100.064303,XIE2023137800,Xie2023}.

To elucidate the underlying mechanism of large MEDs, we conduct a detailed analysis of the radial density distributions of mirror states in $^{18}$Ne/$^{18}$O and $^{19}$Na/$^{19}$O pairs using the GSM. The results allow us to systematically compare the radial distributions of valence protons in proton-rich nuclei and valence neutrons in their neutron-rich mirror counterparts. The results are presented in Figs. \ref{fig-18Ne-18O-density} and \ref{fig-19Na-19O-density} for $^{18}$Ne/$^{18}$O and  $^{19}$Na/$^{19}$O mirror partners, respectively. 
Our GSM results reveal that the states characterized by minor isospin symmetry breaking with small MEDs exhibit almost identical radial density distributions, which decline sharply in the asymptotic regions, such as the ground states of both $^{18}$Ne/$^{18}$O and $^{19}$Na/$^{19}$O. 
This phenomenon is largely attributed to the dominance of $d_{5/2}$ partial waves, which are constrained within the nuclear region by high centrifugal and Coulomb barriers, despite the state being unbound.
Conversely, GSM calculations depict the radial density distributions of the $3_1^+$ state of $^{18}$Ne and the $1/2_1^+$ state of $^{19}$Na as more extended in the asymptotic region than their neutron-rich counterparts, $^{18}$O and $^{19}$O, respectively. This disparity stems from the non-existent centrifugal barrier for the $s_{1/2}$ partial wave, leading to a more pronounced distribution in the proton-rich nucleus due to the weakly bound or unbound nature of the $s_{1/2}$ partial wave. A similar mechanism underlies the formation of halo nuclei, where the valence nucleons occupy weakly bound $s$- or $p$- partial waves, resulting in an extended density distribution due to the minimal or absent centrifugal barrier~\cite{PhysRevC.84.051304,LI2022137225,PhysRevC.101.031301}. Consequently, our GSM calculations provide direct calculations for radial density distribution and unveil that the mirror states demonstrating significant isospin symmetry breaking with large MEDs possess many-body wave functions in proton-rich nuclei that are more extended than those in their neutron-rich mirror states, which add a new dimension to our understanding of the role of isospin symmetry breaking in shaping their properties.

\begin{figure}[!htb]
\includegraphics[width=0.90\columnwidth]{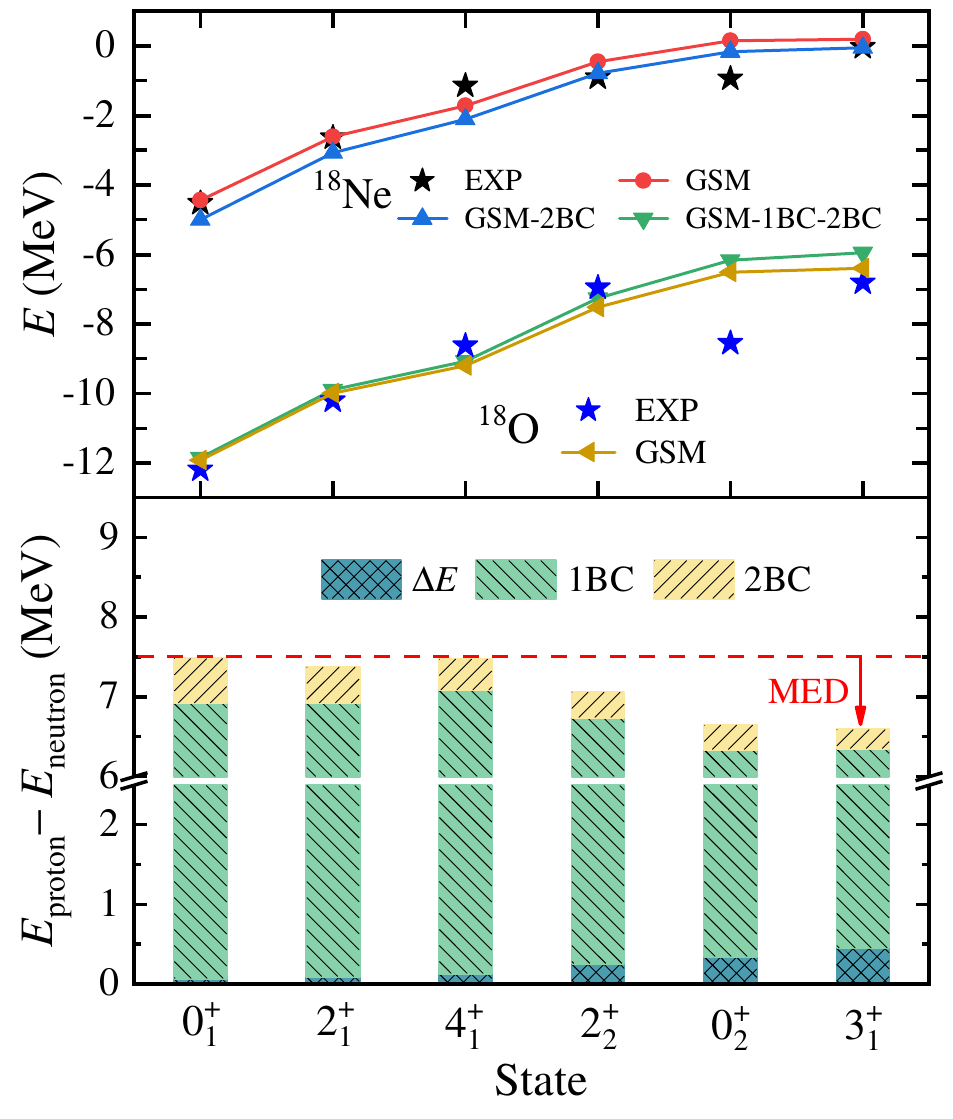}
\caption{Upper panel: calculated energies (GSM), energies minus the two-body Coulomb contribution (GSM-2BC), and energies minus one- and two-body Coulomb total contributions (GSM-1BC-2BC) of low-lying states of $^{18}$Ne using GSM, along with the energies of mirror states in $^{18}$O, with respect to the $^{16}$O inner core. The GSM results are also compared with experimental data. Lower panel: the contribution for the calculated energy difference between the mirror states.}{}\label{18Ne-18O-H-contribution}
\end{figure}

\begin{figure}[!htb]
\includegraphics[width=0.860\columnwidth]{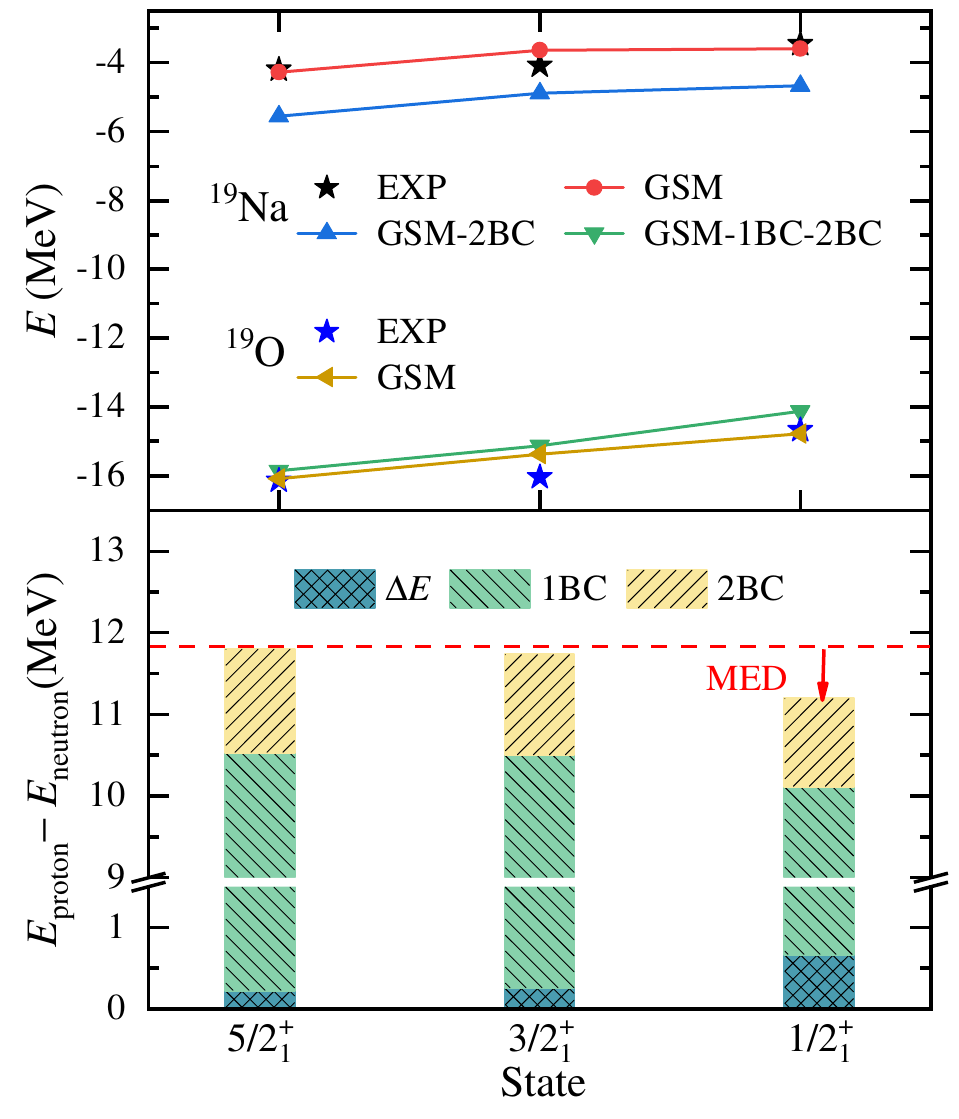}% Here is how to import EPS art
\caption{Similar to Fig.~\ref{18Ne-18O-H-contribution}, but for low-lying states in $^{19}$Na/$^{19}$O.}{}\label{19Na-19O-H-contribution}
\end{figure}

Within the present GSM calculations, the GSM Hamiltonian, as shown in Eq. (\ref{H_GSM}), can be divided into nuclear interaction (encompassing core-nucleons and nucleon-nucleon interaction) and Coulomb interaction (including one-body Coulomb (1BC) interaction between the inner core and valence protons, two-body Coulomb (2BC) interaction between valence protons).
We delve into further calculations to dissect the contributions from different parts of the Hamiltonian, aiming to shed light on the underlying mechanisms in mirror states exhibiting significant isospin symmetry breaking with large MED.

The computed energies for the low-lying mirror states in the pairs $^{18}$Ne/$^{18}$O and $^{19}$Na/$^{19}$O, along with experimental data~\cite{ensdf}, are showcased in Figs.~\ref{18Ne-18O-H-contribution} and \ref{19Na-19O-H-contribution}, respectively. To gain deeper insights, we also present the energy minus 2BC contribution  (GSM-2BC) and energy minus 1BC and 2BC contribution (GSM-1BC-2BC) in proton-rich nuclei $^{18}$Ne and $^{19}$Na.
%The energies, both calculated and experimental, are given with respect to the ground-state energy of $^{16}$O inner core.
Indeed, the GSM-1BC-2BC also corresponds to the contribution of nuclear interaction.
Within the isospin symmetry picture, the difference in mirror state energies should solely stem from Coulomb interactions, implying that the GSM-1BC-2BC values for a state in a proton-rich nucleus would be the same as its mirror state in the neutron-rich nucleus. 

Our GSM calculation gives that the GSM-1BC-2BC values for the ground states of $^{18}$Ne and $^{19}$Na closely align with the computed ground-state energies of their neutron-rich counterparts, $^{18}$O and $^{19}$O, respectively.
The results indicate the preservation of isospin symmetry for these ground states.
Conversely, for the excited $3_1^+$ state in $^{18}$Ne/$^{18}$O and the $1/2_1^+$ state in $^{19}$Na/$^{19}$O, our GSM calculations showcase a deviation from this symmetry. 
Specifically, the GSM-1BC-2BC values for the $3_1^+$ state of $^{18}$Ne and the $1/2_1^+$ state of $^{19}$Na are more unbound compared to their computed energies of mirror states.
To quantitatively examine this discrepancy, we introduce $\Delta E$ as the differential metric. $\Delta E$ encapsulates the disparity between the GSM-1BC-2BC values in the state of the proton-rich nucleus and the energy calculated for the corresponding state in the neutron-rich mirror nucleus, which read as 
$\Delta E = \langle \Psi_{\rm proton} | H_{NN} | \Psi_{\rm proton} \rangle - \langle \Psi_{\rm neutron} | H_{NN} | \Psi_{\rm neutron} \rangle$, %\label{deltaE}
%\end{equation}
the $\Psi_{\rm proton}$ and $\Psi_{\rm neutron}$ correspond to the many-body wave function of proton-rich and neutron-rich nuclei, respectively.
The $\Delta E$ corresponds to the difference in the contribution of nuclear interactions in the mirror state.

Our GSM calculations show that both $\Delta E$ and Coulomb interactions, including 1BC and 2BC , significantly influence the energy discrepancies observed in mirror states. Predominantly, the Coulomb interaction emerges as the dominant factor contributing to these differences. Illustrated in the lower panels of Figs.~\ref{18Ne-18O-H-contribution} and \ref{19Na-19O-H-contribution}, we detail the $\Delta E$, 1BC, and 2BC contributions to the energy differences in the low-lying mirror states of $^{18}$Ne/$^{18}$O and $^{19}$Na/$^{19}$O mirror pairs.
Our GSM results indicate that the energy differences in the ground states of $^{18}$Ne/$^{18}$O primarily stem from Coulomb interactions, with $\Delta E$ making a minimal contribution. Furthermore, for $^{19}$Na/$^{19}$O, the $\Delta E$ contribution is noted to be around 100 keV. Interestingly, we find varying contributions of $\Delta E$, 1BC, and 2BC across different mirror states within each state. For instance, the $3_1^+$ states in $^{18}$Ne/$^{18}$O exhibit a higher $\Delta E$ contribution and lower Coulomb interactions, relative to their ground states.
The heightened $\Delta E$ values underscore the distinct nuclear interaction contributions to isospin symmetry breaking in these systems, showcasing the complex interplay of forces that shape the energy landscapes of mirror nuclei.

In evaluating the MED, the ground state energy of mirror nuclei serves as the baseline, with MED being determined by the discrepancy in excitation energies of corresponding mirror states.
Adopting the energy difference of the ground states of mirror nuclei as a reference---illustrated by red dashed lines in Figs.~\ref{18Ne-18O-H-contribution} and \Ref{19Na-19O-H-contribution}. The difference between the values of ground and excited mirror states corresponds to the MED, highlighted by the red arrows in these figures.
The results reveal that both the $\Delta E$ values and the Coulomb interaction exhibit significant variations across different mirror states, both contributing to the MED.

\textit{Summary.}---Based on the GSM calculations, in which both the inter-nucleon correlation and continuum coupling are properly treated, we deduce that significant isospin symmetry breaking in mirror states, leading to large MED values, arises from the occupation of weakly-bound or unbound $s_{1/2}$ partial waves in the proton-rich nucleus, while its counterpart in the neutron-rich nucleus remains deeply bound. This dichotomy culminates in a more expansive radial density distribution for states within the proton-rich nucleus, as opposed to their mirror counterparts. Additionally, the difference in radial density distributions between mirror states implies disparate contributions from nuclear interactions, underscored by significant $\Delta E$ values, which further highlight the presence of isospin symmetry breaking. Moreover, states with an extended radial density distribution tend to yield smaller Coulomb contributions compared to ground states characterized by more localized distributions. This factor chiefly accounts for the reduced excitation energies in states influenced by the Thomas-Ehrman shift effect, thereby engendering substantial negative MED values in mirror states. Our GSM calculations corroborate that both nuclear and Coulomb interactions play crucial roles in manifesting the significant isospin symmetry breaking associated with significant MED values.

\textit{Acknowledgments} ---
This work has been supported by the National Key R\&D Program of China under Grant No. 2023YFA1606403; the National Natural Science Foundation of China under Grant Nos.  12205340, 12347106, and 12121005;  the Gansu Natural Science Foundation under Grant No. 22JR5RA123 and 23JRRA614; the Key Research Program of the Chinese Academy of Sciences under Grant No. XDPB15; the State Key Laboratory of Nuclear Physics and Technology, Peking University under Grant No. NPT2020KFY13. The numerical calculations in this paper have been done on Hefei advanced computing center.

%\end{acknowledgments}

%%\appendix

\section*{References}

\bibliographystyle{elsarticle-num_noURL}
\bibliography{Ref.bib}% Produces the bibliography via BibTeX.

\end{document}